\documentclass[pra,twocolumn,showpacs,letterpaper,showpacs,superscriptaddress]{revtex4-1}

\usepackage{graphicx,amsmath,amssymb,amsfonts,latexsym,color,dcolumn,bm,epsfig,subfigure}
\usepackage[plainpages=false,hyperfootnotes=false,colorlinks=false]{hyperref}
\usepackage[normalem]{ulem}
\usepackage{dsfont}

\renewcommand{\eqref}[1]{Eq. (\ref{#1})}

\begin{document}

\title{Extended hydrodynamic description for nonequilibrium atom-surface interactions}

\author{D. Reiche}
\affiliation{Humboldt-Universit\"at zu Berlin, Institut f\"ur Physik, AG 
			Theoretische Optik \& Photonik, 12489 Berlin, Germany}
\affiliation{Max-Born-Institut, 12489 Berlin, Germany}

\author{M. Oelschl\"{a}ger}
\affiliation{Humboldt-Universit\"at zu Berlin, Institut f\"ur Physik, AG 
			Theoretische Optik \& Photonik, 12489 Berlin, Germany}
\affiliation{Max-Born-Institut, 12489 Berlin, Germany}

\author{K. Busch}
\affiliation{Humboldt-Universit\"at zu Berlin, Institut f\"ur Physik, AG 
			Theoretische Optik \& Photonik, 12489 Berlin, Germany}
\affiliation{Max-Born-Institut, 12489 Berlin, Germany}

\author{F. Intravaia}
\affiliation{Humboldt-Universit\"at zu Berlin, Institut f\"ur Physik, AG 
			Theoretische Optik \& Photonik, 12489 Berlin, Germany}

% ABSTRACT
%==================================================================================
\begin{abstract}
The dissipative properties of spatially nonlocal conductors are investigated in the context
of quantum friction acting on an atom moving above a macroscopic body. The focus is on an
extended version of the hydrodynamic model for the bulk material's electromagnetic response.
It is shown that the standard hydrodynamic description is inadequate for evaluating the
frictional force since it completely neglects Landau damping. The extended version of the
model contains a frequency-dependent compressibility factor for the Fermi liquid and
qualitatively resolves this issue. For a quantitative assessment, these results are contrasted
with those obtained for the more fundamental Boltzmann-Mermin model. Since the latter is
technically involved, the simplicity of the extended hydrodynamic model allows
for an easier analysis of the impact of nonlocality on quantum friction for other (planar)
geometries. This is illustrated with an example involving a thin slab.
\end{abstract}
%==================================================================================

\renewcommand{\imath}[0]{\mathrm{i}}
\newcommand{\abs}[1]{\left\vert#1\right\vert}
\newcommand{\avg}[1]{\langle#1\rangle}

\newcommand{\mathbfh}[1]{\hat{\mathbf{#1}}}
\newcommand{\mathbft}[1]{\tilde{\mathbf{#1}}}

\newcommand{\arccot}[0]{\text{arccot}}
\newcommand{\arcsinh}[0]{\text{arcsinh}}
\newcommand{\arctanh}[0]{\text{arctanh}}
\newcommand{\arccoth}[0]{\text{arccoth}}

\newcommand{\todo}[1]{{\color{blue} (#1)}}
\newcommand{\new}[1]{{\color{red} #1}}

\maketitle

% INTRODUCTION
%==================================================================================
\section{Introduction}

One remarkable feature of quantum theory is the existence of zero-point fluctuations:
Even in its state of minimal energy each quantum system is restless. These quantum
fluctuations induce macroscopic consequences in the form of experimentally accessible
dispersion forces. One prominent representative of such forces is the Casimir effect
\cite{Casimir48}
which, in its original form, is characterized by an attraction between two electrically
neutral, non-magnetic bodies placed in vacuum.
Over the past decade, another class of quantum-fluctuation-induced phenomena has been
intensively investigated. The accent is placed on the non-conservative aspect of
the interaction. For instance, when spatially separated objects are set in motion relative
to each other, a fluctuation-induced drag force appears and counteracts the dynamics.
This nonequilibrium scenario is usually referred to as quantum or Casimir friction \cite{Pendry97,Dedkov02a,Scheel09,Despoja11,Volokitin11,Silveirinha12,Belen-Farias15}. The frictional interaction strongly depends on the optical
response of the materials composing the objects in the system.
Specifically, in dynamical
nonequilibrium, dispersion forces have shown to be very sensitive to the materials' dissipative characteristics \cite{Volokitin07,dedkov17}.
On the one hand, this means that precise predictions for experimental studies of nonequilibrium
dispersion forces inevitably call for the use of reliable and well-understood models of
the material's optical response. On the other hand, this discloses the potential of these
interactions for usage in (quantum) sensing and other technological applications.

Quantum friction between an atom and a planar surface was investigated in the case of
spatially dispersive conductors in Refs.
\cite{Volokitin03a,Reiche17}.
It was pointed out that, when the atom-surface separation $z_a$ becomes comparable
to the bulk electron's mean free path $\ell$, Landau damping
\cite{Landau46}
plays an important role in determining the strength of the interaction. In addition to
modifying the functional dependence of the force on $z_a$, Landau damping is responsible
for a considerable enhancement of the interaction strength as compared to a local description
that would only feature collision-induced damping (i.e., the Drude model \cite{Jackson75}).
These results have been obtained by describing the reflection coefficients of the metal in the framework provided by the semi-classical infinite
barrier (SCIB) approximation \cite{Ford84}. The dielectric functions where modeled using the Lindhard approach \cite{Lindhard54} including Mermin correction \cite{Mermin70} in the limit of wave vectors smaller than the Fermi wave vector.
Equivalently, one could have used a semi-classical approach relying on the Boltzmann equation and involving the zero-temperature Fermi-Dirac distribution for describing the properties of the electrons in the material. This is why the reported enhancement of the frictional force is sometimes referred to as originating from the \textit{Boltzmann-Mermin model} \cite{Chapuis08}.
Due to its complexity, however, the Boltzmann-Mermin description is not the most practical
approach for describing nonlocal electromagnetic interactions. Numerous studies analyzing
spatial dispersion in nanophotonic systems use the less complex \textit{hydrodynamic model},
which describes the electrons in the medium in terms of a charged and compressible fluid.
Although it might be sufficient for some calculations involving nanostructured objects,
the hydrodynamic model provides a rather rough description of the actual physics of the system. For
example, in its simplest but commonly adopted variant, it does not include Landau damping.
In order to preserve its appealing simplicity without neglecting important physical characteristics,
the hydrodynamic model has been extended in many ways ( see Refs.
\cite{Halevi95,Mortensen14,Toscano15,Moeferdt18}).

In this manuscript, we consider a specific modification of the hydrodynamic equations which
focuses on the compressibility factor $\beta$ of the electronic fluid
\cite{Halevi95} (see Sec. \ref{sec:models}).
Instead of being a real constant as in the standard treatment, the compressibility becomes a
complex function of the frequency of the electromagnetic radiation. In the context of
atom-surface quantum friction (Sec. \ref{sec:friction}), we compare this extended hydrodynamic
model to the Boltzmann-Mermin model on the level of the reflection coefficients for both,
a planar interface to an infinitely extended body (Sec. \ref{sec:halfspace})
and a finite-sized planar slab (Sec. \ref{sec:slab}). Specifically, we
provide a detailed understanding in terms of physical processes and connect our findings
to the material's dissipative processes which for metals coincide with their \textit{resistivity}. Based on this, we study the impact of the two models on quantum frictional forces and assess their range of validity.
We close our discussion with a summary and further remarks (Sec. \ref{sec:concl}).
%==================================================================================

% MODELS
%==================================================================================
\section{Models for spatial dispersion}\label{sec:models}

The description of a spatially nonlocal material and its optical response
at the interface with vacuum or another dielectric has been investigated in various contexts by many authors
(see for example Refs. \cite{Bloch34,Lindhard54,Kliewer68,Jones69,Feibelman82,Mochan83,Esquivel03,Esquivel-Sirvent06,Horovitz12,Teperik13,Mortensen14,Raza15,Schmidt18}).
While in the abstract configuration of an infinite bulk material the symmetry of the system
allows for a self-consistent description of the dynamics, this turns out to be considerably
more difficult as soon as an interface occurs. With a nonlocal description of the material,
additional boundary conditions (ABCs) of non-electromagnetic origin are often required for
a complete characterization of the system's optical response. The two models considered in
our work are no exception. In order to obtain the reflection coefficients for a planar
medium, we follow the SCIB approach which induces an ABC that is equivalent to the specular
reflection of electrons at the interface. The bulk optical response is connected to
the interface impedance by the relations \cite{Ford84}
\begin{subequations}
\label{eq:surfaceimpedanceS}
\begin{gather}
	Z^{\rm TM}(\omega,p)=\frac{2\imath c}{\pi\omega}
		\int_{0}^{\infty}\frac{{\rm d}q}{k^2}
		\left[\frac{q^2}{\epsilon_t(\omega,k)-\frac{c^{2}k^2}{\omega^{2}}}
			+\frac{p^2}{\epsilon_l(\omega,k)}
		\right],
\label{eq:surfaceimpedanceP}\\
	Z^{\rm TE}(\omega,p)=\frac{2\imath c}{\pi\omega}
		\int_{0}^{\infty}{\rm d}q~
		\frac{1}{\epsilon_t(\omega,k)-\frac{c^{2}k^2}{\omega^{2}}},
\end{gather}
\end{subequations}
where $c$ is the vacuum speed of light and TE and TM describe the two polarizations of the
electromagnetic radiation.  $\epsilon_l(\omega,k)$ and  $\epsilon_t(\omega,k)$
denote the nonlocal longitudinal and transverse (bulk) permittivities. They are functions
of frequency $\omega$ and, due to symmetry reasons, of the modulus of three-dimensional wave vector
$k^{2}=p^{2}+q^{2}$, where $p=\sqrt{p_{x}^{2}+p_{y}^{2}}$ is the component parallel to the
surface and $q$ is the component orthogonal to it.

In previous work
\cite{Reiche17},
we have used these expressions in conjunction with the permittivity functions given by the
Boltzmann-Mermin model for a metal
\cite{Lindhard54,Mermin70,Ford84},
which we report here for completeness:
\begin{subequations}
\label{eq:permittivitiesL}
 \begin{align}
	\epsilon_l(\omega,k)
		&=1+\frac{\omega_\mathrm{p}^2}{\omega+\imath\Gamma}
			\frac{3u^2 g_l(u)}{\omega+\imath\Gamma g_l(u)},\\
	\epsilon_t(\omega,k)
		&=1-\frac{\omega_\mathrm{p}^2}{\omega(\omega+\imath\Gamma)}g_t(u),
 \end{align}
 \end{subequations}
where $\omega_{\rm p}$ is the plasma frequency of the metal and
\begin{subequations}
\label{gfunct}
 \begin{gather}
  g_l(u)=1-u \, \arccoth\left(u\right)\\
  g_t(u)=\frac{3}{2}\left[u^{2}-(u^2-1)u \, \arccoth\left(u\right)\right]
 \end{gather}
\end{subequations}
are dimensionless functions of the variable $u=(\omega+\imath\Gamma)/(v_{\rm F} k)$ defined
in terms of the Fermi velocity $v_{F}$ and the collision (dissipation) rate $\Gamma$.
For simplicity, we are going to consider a constant value of $\Gamma$ in the following. In general, however, $\Gamma$ contains a temperature-dependent part connected with the type of collision (e.g. electron-electron, electron-phonon) \cite{bloch30,grueneisen33}, as well as a temperature-independent part depending on the amount of defects or impurities present in the material \cite{bass90,footnote1}. Eqs. (\ref{eq:permittivitiesL}) and (\ref{gfunct}) are also consistent with the assumption of a fully degenerated electron gas, valid for $T\ll T_{\rm F}$ ($T_{\rm F} \sim 10^5$ K).
Based on the Lindhard framework
\cite{Lindhard54},
\eqref{eq:permittivitiesL} includes the so-called Mermin correction
\cite{Mermin70,Ford84},
but neglects effects occurring at scales below the electron's de Broglie wavelength, $\lambda_{\rm B}=2\pi \hbar/(m v_{\rm F})$ (essentially the inverse of the  Fermi wave vector and usually
of the order of a few tenths of an angstrom), which are outside the range of validity of the theory. The
local limit is obtained for $\abs{u}\to \infty$, where both longitudinal and transverse
permittivity coincide with the well-known spatially local Drude permittivity
\begin{align}\label{eq:Dr}
 \epsilon_{\rm Dr}=1-\frac{\omega_\mathrm{p}^2}{\omega(\omega+\imath\Gamma)}.
\end{align}
The strongest nonlocal corrections, however, occur for $\abs{u}\to 0$: The dielectric functions show a nonzero complex part connected with the
(mathematical) behavior of $\arccoth[u]$ for $u<1$ even if no collision-induced damping
is present ($\Gamma\to 0$). Physically, the complex part corresponds to Landau damping.
This phenomenon describes a situation where the phase velocity of the electromagnetic wave in the electron gas matches or is less
than the electrons' speed, i.e. $\omega/k \le v_{F}$  allowing for the transfer of the electromagnetic field’s energy to the electrons.

As explained above, the focus of our work lies on the so-called hydrodynamic model. It derives from a description of the electrons in the bulk as quasi-particle excitations
of a charged and compressible fluid. The quasi-particle excitations are supposed to obey both, charge conservation (continuity equation) and momentum conservation. The latter is described
by means of the Euler equation including collision-induced damping and the Fermi pressure
(originating from the Pauli exclusion principle) under the influence of the external Lorentz
force \cite{huynh18}.
In its linearized version, the permittivity functions for the hydrodynamic model are given
by the rather simple expressions
\begin{subequations}
\label{eq:permittivitiesH}
 \begin{align}
	\epsilon_l(\omega,k)
		&=1-\frac{\omega_\mathrm{p}^2}{\omega(\omega+\imath\Gamma)-\beta^{2}k^{2}},\\
	\epsilon_t(\omega,k)
		&=1-\frac{\omega_\mathrm{p}^2}{\omega(\omega+\imath\Gamma)}.
 \end{align}
 \end{subequations}
Notice that in this description only the longitudinal permittivity is affected by nonlocality,
while the transverse one is identical to the common local expression given by the Drude model,
i.e. $\epsilon_t(\omega,k)\equiv \epsilon_{\rm Dr}(\omega)$ \cite{Jackson75}. In contrast to the Boltzmann-Mermin model, only the TM polarized reflection coefficient is
modified relative to its local counterpart [see \eqref{eq:surfaceimpedanceS}]. The most important
parameter of the hydrodynamic description is the compressibility of the electron fluid, described
by $\beta$, which in our case is connected to the Fermi pressure
\cite{Halevi95}.
In the literature, authors have chosen different values for this parameter, the most popular
being $\beta^{2}=v^{2}_{F}/3$ and $\beta^{2}=3v^{2}_{F}/5$. They correspond to different frequency
limits (low frequency -- Thomas-Fermi description
\cite{Kittel96}
and high frequency, respectively) in the implicit assumptions which lead to the hydrodynamic
equations of motion of a degenerate electron gas (for instance,
see Refs.
\cite{Barton79,Halevi95,Atwal02}).
Independent of its precise value, however, as long as $\beta$ stays real, the imaginary part of
the permittivity is intrinsically connected with the collision-induced damping $\Gamma$ only.
For $\Gamma=0$, apart from a resonant exchange of energy occurring at $\omega/k=\beta$, the
dielectric function describes a system lacking dissipation.

With the aim of providing a more fundamental background for the hydrodynamic description, P. Halevi
suggested to compare the two models, i.e. the Boltzmann-Mermin and hydrodynamic model
\cite{Halevi95}.
In particular, by equating the two longitudinal permittivities in \eqref{eq:permittivitiesL}
and \eqref{eq:permittivitiesH}, we obtain
\begin{equation}
 \frac{\beta^{2}}{v_{F}^{2}}=
 \frac{\omega\left[u^{2}+\frac{1}{3g_{l}(u)}\right]+\imath\frac{\Gamma}{3}}{
      \omega +\imath\Gamma}.
\end{equation}
An expansion in $1/\abs{u}$ gives at the leading order the complex and frequency-dependent
compressibility factor,
\begin{equation}
 \beta^{2}(\omega)\approx
 \frac{\frac{3}{5}\omega+ \frac{\imath}{3}\Gamma}{
 \omega+\imath \Gamma}v_{F}^{2}.
 \label{eq:betaHalevi}
\end{equation}
This function introduces an additional phase between the plasma oscillations and the
electromagnetic radiation. It also adequately reproduces the limiting values of $\beta$
for the low and and high frequency limit, respectively.
Interestingly enough, despite the fact that the expansion in $1/\abs{u}$ is consistent
with a limit of weak spatial dispersion, some features of the Landau damping are preserved
as long as the collisional damping is large enough. Indeed from $1/\abs{u}\ll 1$, we obtain
\begin{equation}
 \sqrt{v^{2}_{F}-\frac{\Gamma^{2}}{k^{2}}}\ll \frac{\omega}{k}\le v_{F},
\end{equation}
which sets a lower bound to the validity of the approximation in terms of the phase velocity
of the electromagnetic field. More importantly, it shows that the approximation
(\ref{eq:betaHalevi}) still allows for waves with a phase velocity smaller than the
electrons' speed, hence allowing for Landau damping.
Since Landau damping plays a significant role in quantum friction at small atom-surface separations, it is interesting to
analyze the impact of \eqref{eq:betaHalevi} on the drag force. We study whether the extension
to the usual hydrodynamic model proposed by Halevi can reproduce the features of
the more fundamental Boltzmann-Mermin model.
Finally, we like to note that recently
Mortensen et al.
\cite{Mortensen14}
have suggested a similar extension of the standard hydrodynamic model via a complex
compressibility coefficient in order to capture the physics of nano-gap
structures in numerical evaluations.
%==================================================================================

% QUANTUM FRICTION
%==================================================================================
\section{Atom-surface quantum friction}\label{sec:friction}

For our purpose, it is convenient to briefly review the theory of quantum
friction. We refer to the literature for more details (for instance, see Refs.
\cite{Volokitin02,Intravaia16a,dedkov17}).
Throughout the manuscript we assume for simplicity that system is at zero temperature.
For an atom or another microscopic object (nano-particle) moving in vacuum at constant
velocity $\mathbf{v}$ and constant height $z_a$ above a flat surface,
the quantum frictional force $\mathbf{F}$ is given by
~\cite{Intravaia14,Intravaia16a}
\begin{multline}
  \mathbf{F}
    =
  -2\int_0^\infty \mathrm{d}\omega\,\int \frac{\mathrm{d}^2\mathbf{p}}{(2\pi)^2} \\
	\times \mathbf{p}\, \mathrm{Tr}\left[
                                    \underline{S}(\mathbf{p}\cdot\mathbf{v}-\omega,\mathbf{v})
                                    \cdot
                                    \underline{G}_I(\mathbf{p},z_a,\omega)
                                 \right] \,.
  \label{eq:QFint}
\end{multline}
The subscript $I$ indicates the imaginary part of the corresponding quantity. $\underline{S}(\omega, \mathbf{v})$ is the velocity-dependent atomic power spectrum and
$\underline{G}$ is the Fourier transform with respect to the planar coordinates of
the electromagnetic Green tensor. For a flat surface, the expression of the latter
can be given in terms of a linear combination of the transverse electric and magnetic
reflection coefficients
\cite{Wylie84}
[see Eqs. (\ref{eq:reflectioncoefficients}) and (\ref{eq:refslab})].
For simplicity, in \eqref{eq:QFint} we model the moving particle as a rigid dipole
$\hat{\mathbf{d}}(t)=\mathbf{d}\hat{q}(t)$, where $\mathbf{d}$ is the static dipole
vector and $\hat{q}(t)$ describes the dipole's internal dynamics.
Although quantum friction shares many aspects with the equilibrium Casimir-Polder
interaction, it clearly displays features which are specific of a non-conservative
interaction. Specfically, the drag force is strongly connected with the system's
dissipative behavior. In our setup, dissipation can arise in (i) the macroscopic
material of the surface and (ii) the particle's internal degrees of freedom.
For the metallic surface (i), this points to the collision-induced damping and, if
present, to Landau damping of the electron fluid. In the case of the moving microscopic
object (ii), the origin of dissipation is different depending on whether the particle
features some form of internal dissipation (as it is the case, e.g., for a metallic
nanoparticle) or whether the damping is induced by the interaction with the electromagnetic
field only, as it is the case in an atomic system. In this work, we only consider the
latter case, where the radiation dressing, backaction and their interplay with the
surface characterize the interaction between atom and surface
\cite{Intravaia16}.
If the internal dynamics is described in terms of a harmonic oscillator, the atomic
power spectrum reads as
\begin{multline}
 \underline{S}(\omega,\mathbf{v})=
 \frac{\hbar}{\pi}\int \frac{\mathrm{d}^2\mathbft{p}}{(2\pi)^2}\theta(\omega+\mathbft{p}\cdot\mathbf{v})\\
 \times\underline{\alpha}(\omega,\mathbf{v})\cdot\underline{G}_I(\mathbft{p},z_{a},\omega+\mathbft{p}\cdot\mathbf{v})
 \cdot\underline{\alpha}^*(\omega,\mathbf{v}),
\end{multline}
where $\underline{\alpha}(\omega,\mathbf{v})$ is the velocity-dependent atomic polarizability and $\theta(\omega)$ is the Heaviside step function \cite{Intravaia16}.

In our non-relativistic description, quantum friction on atoms originates from the interaction
at frequencies $0\le\omega\lesssim v/z_{a}$ ($v=|\mathbf{v}|$). Hence,  for sufficiently low
velocities, only the low-frequency features of the atomic power spectrum and the bulk material's
optical response are of interest. For many practical cases, it is therefore sufficient to restrict
the description of materials to their conducting properties.
For typical velocities (speed of sound and slower), this corresponds to a frequency region where
common materials (including those considered in our work) are Ohmic, i.e. $r_I(\omega,k) \propto \omega$.
Under these circumstances and for a motion within the near field of the surface (strongest interaction), the force is well approximated by~\cite{Reiche17}
\begin{align}
 F &\approx- 2\hbar\frac{v^{3}}{\pi}\left(\frac{\Phi_{0}\Phi_{2}}{3}\mathcal{D}_{0}(z_{a})\mathcal{D}_{2}(z_{a})+\Phi_{1}^{2}\mathcal{D}^{2}_{1}(z_{a})\right).
 \label{eq:generalAsymp}
\end{align}
Here, $\mathbf{F}=F\mathbf{v}/v$ and we have defined
\begin{subequations}
 \begin{align}
  \Phi_{n} &= \binom{2n}{n}\frac{\frac{2n+1}{2(n+1)} \alpha_{xx} + \frac{1}{2(n+1)} \alpha_{yy} + \alpha_{zz}}{2^{2n+3}\pi\epsilon_0}~,
  \label{eq:Geometry}\\
  \mathcal{D}_{n}(z_{a})&=\int_{0}^{\infty}\mathrm{d}p\,
  p^{2 (n+1)}e^{-2z_{a}p}[\partial_{\omega}r^{\rm TM}_{I}(\omega,p)_{\vert \omega=0}],
  \label{eq:DistanceDep}
 \end{align}
\end{subequations}
where $\epsilon_{0}$ denotes the vacuum permittivity.
The constants $\alpha_{ii}$ are the diagonal components of the static atomic polarizability tensor,
which for our model is given by the dyadic $\underline{\alpha}_{0}=2\mathbf{d}\mathbf{d}/\hbar \omega_{a}$
with $\omega_{a}$ being the characteristic atomic transition frequency.
The sign in \eqref{eq:generalAsymp} highlights that the force is oriented opposite to the direction
of motion and, as expected, the definition of $\mathcal{D}_{n}(z_{a})$ shows that it is connected to
the dissipative properties of the surface (imaginary part of the reflection coefficient).
Notice that in the near-field limit, as long as the atom is described in terms of an electric dipole, the dominant contribution to the interaction is provided solely by the TM-polarized field.
Further,
$\mathcal{D}_{n}(z_{a})$ also defines the functional dependence of the frictional force on the
atom-surface separation $z_{a}$. For a spatially local material, the reflection coefficients are
independent of the wave vector and we find in the Ohmic regime
$[\partial_{\omega}r^{\rm TM}_{I}(\omega)_{\vert \omega=0}]=2\epsilon_{0}\rho_{\rm lc}$, where
$\rho_{\rm lc}$ is a constant related to dissipation in the material. For the local Drude conductor, the permittivity at low frequencies is asymptotic to $\epsilon\sim\imath[\epsilon_0\rho_{\rm lc}\omega]^{-1}$, where $\rho_{\rm lc}=\Gamma/(\epsilon_0\omega_\mathrm{p}^2)$ is the metal's resistivity.
The integral in \eqref{eq:DistanceDep} can be readily evaluated to
\begin{equation}
\mathcal{D}_{n}(z_{a})=2\epsilon_{0}\rho_{\rm lc} \frac{[2(n+1)]!}{(2z_{a})^{2n+3}}
\label{D_definition}
\end{equation}
demonstrating that $F_{\rm lc}\propto \rho_{\rm lc}^{2}v^{3}/z_{a}^{10}$ \cite{Reiche17}.
%==================================================================================

% BULK
%==================================================================================
\begin{figure}
 \centering
 \includegraphics[width=\linewidth]{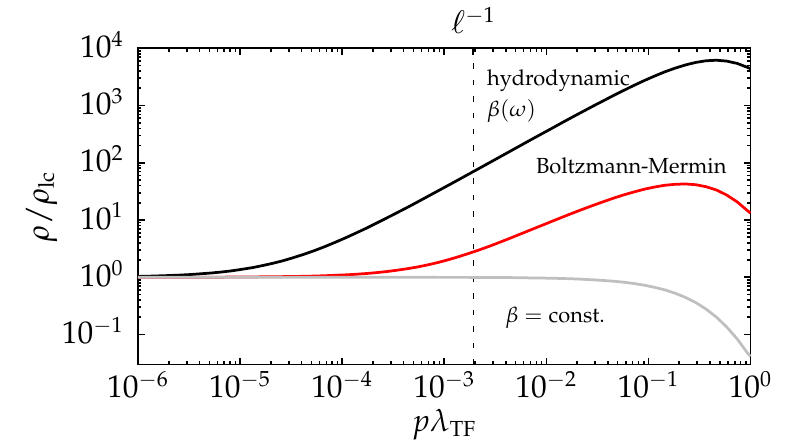}
  \caption{The nonlocal resistivity $\rho(p)$ as a function of the wave vector $p$ and normalized
	         to its local counterpart, $\rho_{\rm lc}$, for the different models analyzed in the main
					 text. Both, the Boltzmann-Mermin (red) and the extended hydrodynamic model with a
					 frequency-dependent compressibility factor (black) feature a maximum in $\rho(p)$ for
					 $p\lambda_{\rm TF}\sim 1/5$ and $p\lambda_{\rm TF}\sim 1/2$, respectively. The standard
					 hydrodynamic description (gray) with constant and real $\beta$ (here equal to $v_{F}/\sqrt{3}$)
					 corresponds to a $\rho(p)/\rho_{\rm lc}$ which tends to one for $p\lambda_\mathrm{TF}\ll 1$ and
					 monotonically decreases to zero for $p\lambda_\mathrm{TF}\gg 1$.
					 Typical parameters for gold have been chosen, namely $\Gamma=30$ meV, $\omega_\mathrm{p}=9$ eV
					 \cite{barchiesi2014}
					 and $v_\mathrm{F}=\alpha_{\rm fs} c$ with $\alpha_{\rm fs}$ the fine-structure constant. }
\label{fig:rho}
\end{figure}
\subsection{A nonlocal infinite half space}\label{sec:halfspace}
The above expressions also indicate how spatial dispersion can modify the frictional interaction.
While the dependence on the velocity is essentially not affected by nonlocality, the force can
be modified both, in its strength and in its functional dependence on the atom-surface separation
through the expression for $\mathcal{D}_{n}(z_{a})$.
In simple terms, the main difference with respect to the local case is that the material resistivity
is no longer described by a constant. Rather, as soon as spatial nonlocality is considered, the
material resistivity becomes a function of the wave vector parallel to the surface $p$. In analogy
with the local description we define the nonlocal resistivity as
$\rho(p)\equiv \partial_{\omega}r^{\rm TM}_{I}(\omega,p)_{\vert \omega=0}/(2\epsilon_{0})$. In
principle, one could also define a resistivity  connected to the transverse electric reflection
coefficient. Since the TE-waves give a sub-leading contribution in the near field, we skipped this distinction in order
to keep the notational complexity at bay.
The TM-reflection coefficient, for the interface to an infinitely extended half space medium can
be written as a function of the surface impedance, i.e.
\begin{align}\label{eq:reflectioncoefficients}
 r^{\rm TM}(\omega,p)=\frac{1-Z^{\rm TM}/Z^{\rm TM}_0}{1+Z^{\rm TM}/Z^{\rm TM}_0},
% =r^{\rm TM}(0,p)+2\imath\epsilon_0\rho(p)\omega+\mathcal{O}(\omega^2),
\end{align}
where $Z_0^{\rm TM}$ is the corresponding relation of the vacuum given by Eq. (\ref{eq:surfaceimpedanceP})
with $\epsilon=1$. We obtain for the spatially nonlocal resistivity
\begin{align}\label{eq:rhononlocal}
\rho(p) &=
 \frac{2}{\pi \epsilon_{0}}
 \frac{\int_{0}^{\infty}\mathrm{d}q~\frac{p}{k^{2}}\mathrm{Im}\left\{\frac{\partial_{\omega}\epsilon_l(\omega,k)_{\vert \omega=0}}{[\epsilon_l(0,k)]^{2}}\right\}}{\left[1+\frac{2}{\pi}\int_{0}^{\infty}\mathrm{d}q~\frac{p}{k^{2}\epsilon_l(0,k)}\right]^{2}}.
\end{align}
In general, \eqref{eq:rhononlocal} depends on all the intrinsic length-scales
characterizing the material. The most relevant ones for our analysis are the Thomas-Fermi screening
length $\lambda_{\rm TF}=v_{F}/(\sqrt{3}\omega_{p})$ and the mean free path $\ell =v_{F}/\Gamma$,
with typical values between a few angstroms and a few tens of nanometers, respectively.
Figure \ref{fig:rho} shows the resistivity as a function of the wavector for the permittivities
in \eqref{eq:permittivitiesL} and \eqref{eq:permittivitiesH}.
For the description based on the Boltzmann-Mermin model, the resistivity $\rho(p)$ features a
maximum around $p\sim 1/(5 \lambda_{\rm TF})$, where it can be more than an order of magnitude
larger than its local counterpart, $\rho_{\rm lc}=\Gamma/(\epsilon_{0}\omega_\mathrm{p}^{2})$.
In the case of the hydrodynamic description with a frequency-dependent compressibility factor,
the resistivity has instead a maximum for $p\sim 1/(2 \lambda_{\rm TF})$ with a value that is
$\sim 2\omega_\mathrm{p}^{2}/(5\Gamma^{2})$ times larger than the value for the local description.
At this point, we should like to note that the behavior of $\beta(\omega)$ is essential for this
result: The resistivity for the extended hydrodynamic model can indeed be written as the sum of
two contributions
\begin{align}\label{eq:rhosplit}
\rho_{\rm eH}(p)=\rho_{\rm H}(p)+\rho_{\rm Ld}(p).
\end{align}
The expression $\rho_{\rm H}(p)$ is the resistivity one would have obtained for a constant value of $\beta=\beta(0)$, i.e. for the
usual common hydrodynamic description, while $\rho_{\rm Ld}(p)$ is
a correction due to the frequency dependence of the compressibility factor ($\sim\partial_{\omega}\beta(\omega)_{|\omega=0}$).
The first term is dominated by the collision-induced damping of
quasi-particles in the conductor. It monotonously decreases from the value $\rho_{\rm lc}$ for
$p\lambda_\mathrm{TF}\ll 1$ to zero for $p\lambda_\mathrm{TF}\gg 1$.
It is, therefore, the second contribution, $\rho_{\rm Ld}(p)$, which dominates $\rho_{\rm eH}(p)$ in
the nonlocal regime and gives rise to the aforementioned maximum. Due to its intrinsic connection to the frequency-dependence of the compressibility factor $\beta(\omega)$, $\rho_{\rm Ld}(p)$ physically encodes
the process of Landau damping. For very large wave vectors, $p\gg\ell^{-1}$, spatial dispersion
becomes negligible and both the hydrodynamic and the Boltzmann-Mermin model (see Fig. \ref{fig:rho}
and Ref. \cite{Reiche17}) approach the local Drude description. More detailed expressions can be
found in the Appendix \ref{appendix1}. %\ref{appendix1} and \ref{appendix2}.
\begin{figure}
 \centering
 \includegraphics[width=\linewidth]{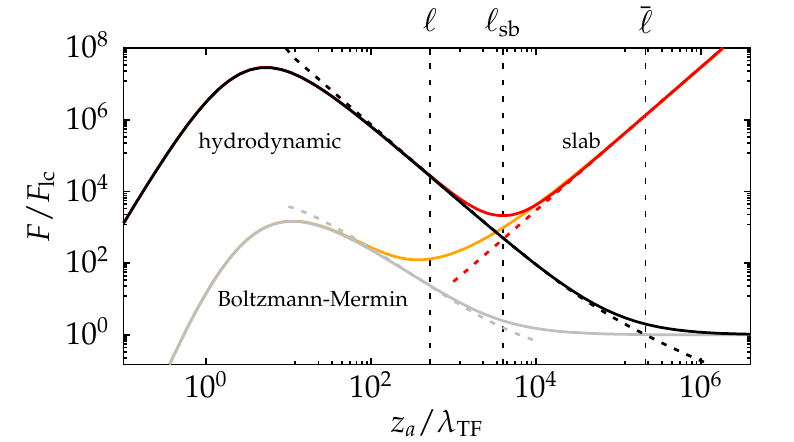}
 \caption{Quantum frictional force experienced by an atom moving at low velocity parallel to a planar interface
          at separation $z_a$ [\eqref{eq:generalAsymp}].
          We normalize to the
		  local Drude result $F_{\rm lc}$ (see Eq. (\ref{D_definition})
		  and Ref. \cite{Reiche17}).
		  The drag force near a semi-infinite bulk and a slab with thickness $d=10^2\lambda_{\rm TF} \approx 9 nm$ is depicted. The black (bulk) and the red (slab) lines corresponds to the permittivity given by the extended hydrodynamic model.
The gray (bulk) and the orange (slab) lines refer instead to the Boltzmann-Mermin model. The asymptotes of Eqs. (\ref{eq:fricbulkhydro}) and (\ref{eq:asympslab}) as well as for the Boltzmann-Mermin model from Ref. \cite{Reiche17} are given in dashed lines. Parameters are chosen as in  Fig.\ref{fig:rho}. }
 \label{fig:ForceDistanceDep}
\end{figure}
Inspecting the force in \eqref{eq:QFint} and the definition of $\mathcal{D}_{n}(z_{a})$, we can
conclude that a \textit{constant} compressibility factor in the (nonlocal) hydrodynamic description modifies
the quantum frictional force relative to that with the local counterpart of the material model only
for distances $z_{a}\ll \lambda_{\rm TF}$. These are distances outside the range of validity of our
theory. Conversely,  if $\beta=\beta(\omega)$, modifications already occur for considerably larger
atom-surface separations ($z_a\sim\ell$). Figure \ref{fig:rho} shows that, although the resistivity of the extended hydrodynamic model qualitatively
resembles the Boltzmann-Mermin model, it quantitatively overestimates the impact of nonlocality.

\begin{figure}
% \begin{center}
 \includegraphics[width=\linewidth]{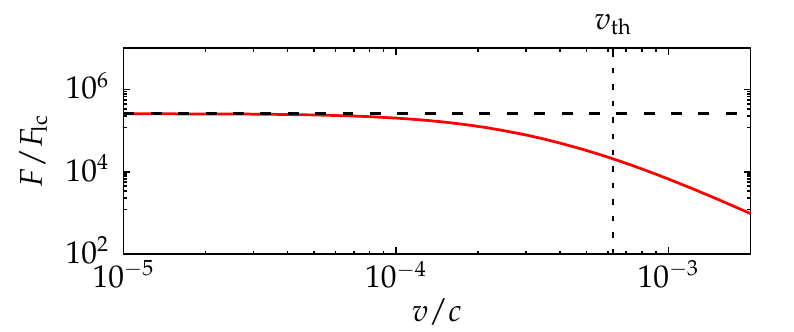}
% \end{center}
  \caption{Full retarded velocity dependence of the quantum frictional force using the extended hydrodynamic model
	         for a half-space geometry.
	         The force is normalized by the non-retarded Ohmic result obtained for the local Drude model [see after \eqref{D_definition}]
	         to highlight the deviation from the $v^{3}$ behavior.
	         The atom-surface separation
					 was chosen to be $z_a=10\,\mathrm{nm}\sim 10^{2}\lambda_{\rm TF}$, where the standard
					 hydrodynamic model with constant
					 and real $\beta$ and the local Drude model  coincide. The parameters are chosen as
					 in Fig.~\ref{fig:rho}. The value $v_\mathrm{th}$ [see \eqref{eq:vth}] is an estimate for
					 the breakdown of the Ohmic regime. The dashed
					 horizontal line represents the Ohmic approximation of \eqref{eq:fricbulkhydro}.\label{fig:vplot}}
\end{figure}

This behavior is reflected in the frictional force. Indeed, from \eqref{eq:DistanceDep} we obtain that the
dominant contributions to the force in the planar setup stem from wave vectors $p\sim 1/(2 z_{a})$.
In accordance with the maxima of the corresponding resistivity (see Fig. \ref{fig:rho}), quantum friction
thus  features a maximum at $z_{a}\sim 10 \lambda_{\rm TF}$ for the Boltzmann-Mermin description
and a maximum at $z_{a}\sim 4 \lambda_{\rm TF}$ for the hydrodynamic model. Further, both models
predict an enhancement of several orders of magnitude for the frictional interaction. Again, despite the
qualitative similarities in the behavior of the force, if the Boltzmann-Mermin description is taken
as reference, the extended hydrodynamic model gives rise to a force which is several orders of magnitude
larger (see Fig. \ref{fig:ForceDistanceDep}). To be more concrete, we average over all dipole orientations in Eq. (\ref{eq:generalAsymp})
and find that in the nonlocal regime ($z_a\sim\ell$)
\begin{align}\label{eq:fricbulkhydro}
\frac{F}{F_{\rm lc}}\approx\frac{\omega_{\rm p}^2}{\Gamma^2}
\left(
2\frac{\lambda_{\rm TF}}{z_a}+\frac{73}{240}\frac{\ell^2}{z_a^2}
\right).
\end{align}
For separations $z_a\gg\bar{\ell}\approx 1.4 \frac{\omega_\mathrm{p}}{\Gamma}\ell$, Landau damping completely
looses impact and the extended hydrodynamic description transitions into the regime dominated by
spatially local (Drude) physics (see Appendix \ref{appendix1}).
Employing the extended hydrodynamic description, the frictional force is still described by a power
law, but as compared to the local model with a different exponent. In contrast, in Ref. \cite{Reiche17}
it was shown that the Boltzmann-Mermin model produces a logarithmic behavior. Still, the hydrodynamic
model for conducting materials in the form suggested by Halevi offers an analytically much simpler
structure whilst conserving the most important physical features. In this way, it becomes adequate
for qualitatively studying more complicated geometries which we illustrate in the following section by considering
the interaction with a planar slab (thin film) of metal.
\begin{figure}
 \centering
 \includegraphics[width=\linewidth]{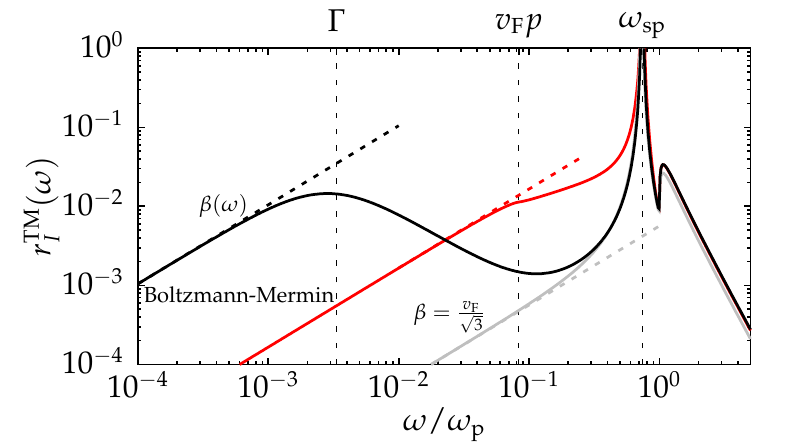}
 \caption{Frequency dependence of the imaginary part of the TM reflection coefficient of a planar
          interface between vacuum and metal-filled half-space. The permittivity of the bulk is modeled
					using the Boltzmann-Mermin model (red), the extended hydrodynamic description (black)
					and standard hydrodynamic model (gray).
					The linear (Ohmic) regime is given in dashed lines.
					The parameters are chosen as in Fig. \ref{fig:rho} and $p=50/(2\ell)$.
					For these values the local (Drude) model (not shown) is almost identical to the standard hydrodynamic description.
\label{fig:reflectionCoeff}}
\end{figure}
At this point, it is interesting to discuss the validity of the Ohmic approximation for the
material models considerd in our work (see Fig. \ref{fig:reflectionCoeff}). While all descriptions
are essentially equivalent for frequencies larger or of the order of the surface plasmon-polariton
resonance $\omega_{\rm sp}$ (see Ref.
\cite{pitarke07}),
they are  very different in the low frequency region. The Ohmic range, where the $r^{\rm TM}_{I} \propto \omega$,
is characteristic of each model.
While for the local description and the standard hydrodynamic expression $r^{\rm TM}_{I} \propto \omega$
for $\omega\ll \omega_{\rm sp}$, the Boltzmann-Mermin model as well as the extended hydrodynamic model feature additional length scales limiting the Ohmic
approximation. The Boltzmann-Mermin model displays Ohmic behavior as long as the radiation's frequency
dominating the interaction $\omega$ is smaller than either the collision-rate $\Gamma$ or than the
value $v_\mathrm{F} p$. In other words, the material is Ohmic as long as some damping mechanism (either
collision-induced or Landau damping) is present. Note that both frequency scales are encoded in the
parameter $u$ [see its definition after Eqs. (\ref{gfunct})]. In the extended hydrodynamic model with the Halevi correction, however,
only one of the above frequency scales is present.
Mathematically, this can be seen from the linear relation between the compressibility factor and the
Fermi velocity $\beta\propto v_\mathrm{F}(\frac{3}{5}\omega+\frac{\imath}{3}\Gamma)/(\omega+\imath\Gamma)$.
Since $v_\mathrm{F}$ is constant, the only intrinsic frequency scale derives from the complex prefactor and
is given by the dissipation rate $\Gamma$. For frequencies $\omega\gg\Gamma$, the model automatically
approaches its high-frequency limit.
Further, we want to comment on the range of validity of the asymptotic expression in \eqref{eq:generalAsymp} which is based on the Ohmic behavior of the reflection coefficient.
Since the scaling of the reflection coefficient with frequency is intrinsically connected to the
velocity scaling of the force
\cite{Oelschlager18},
each model leads to a different velocity dependence of the frictional force above the Ohmic threshold.
As a rough estimate of the threshold velocity $v_{\rm th}$ for the hydrodynamic model with frequency-dependent
$\beta$, we compare its linear regime with the value of the reflection coefficient at its plateau,
where $\omega\sim\Gamma$ (see black curve in Fig. \ref{fig:reflectionCoeff}). This constraints the
validity of the linear regime of the extended hydrodynamic model to frequencies
$\omega\lesssim r^{\rm{TM}}_I(\Gamma,p)/(2\epsilon_0\rho_{\rm eH}(p))$ ($p\ll \lambda_{\rm TF}^{-1}$),
where $\rho_{\rm eH}$ is the respective resistivity. Since the frictional force on atoms is dominated
by frequencies $\omega\lesssim \mathbf{p}\cdot\mathbf{v}$ and for the planar geometry we have that
$p\sim z_a^{-1}$ \cite{Intravaia16}, we conclude that the Ohmic response of the extended hydrodynamic
model holds for velocities (see Fig. \ref{fig:vplot})

\begin{align}\label{eq:vth}
v\ll v_{\rm th}=\frac{z_a}{2\epsilon_0\rho_{\rm eH}}r_I^{\rm TM}(\Gamma,z_a^{-1}).
\end{align}
Notice that, due to the integration over the frequency appearing in the definition of quantum friction,
these features are smoothed out in the force, reducing these
differences in the functional dependence on the velocity as can be seen in Fig.~\ref{fig:vplot}.

%

%%
%%

%

%==================================================================================

% SLAB
%==================================================================================
\subsection{Nonlocal thin films}\label{sec:slab}

\begin{figure}
 \includegraphics[width=\linewidth]{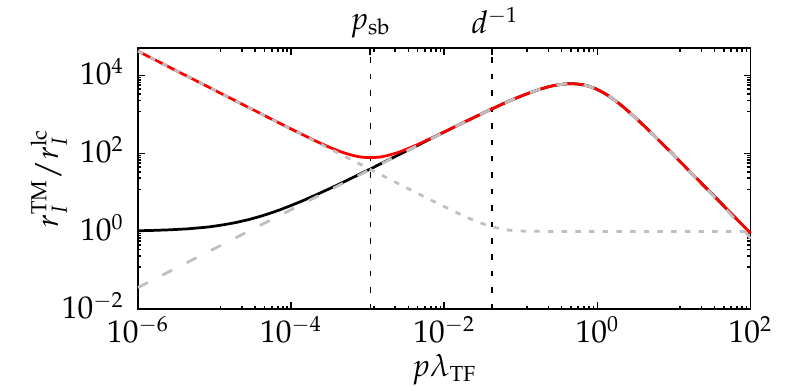}
 \caption{Imaginary part of the TM-reflection coefficient normalized to the corresponding low-frequency Drude result,
					$r^\mathrm{lc}_I\approx 2\epsilon_0\rho_\mathrm{lc}\omega $, as a function of the wave vector $p$ for different
 					geometries and given frequency $\omega/\omega_\mathrm{p}=10^{-5}$ (Ohmic region).
					The slab's (red solid line) and the half-space's reflection
					coefficient (black solid line)
					feature a different behavior for small wave vectors $p\ll p_\mathrm{sb}$ [see
					\eqref{eq:rhoslab} and below].
					We report the corresponding asymptotic characteristics
					in gray (dashed line for $\rho_{\rm Ld}(p)$ and dotted line for $\rho_\mathrm{H}(p)\coth[dp]$).
					The parameters are the same as in Fig.~\ref{fig:rho}
					and we set the layer thickness
					to $d=1\,\mathrm{nm}\sim 10 \lambda_{\rm TF}$.
\label{fig:refk}}
\end{figure}

One major (technical) inconvenience of the Boltzmann-Mermin model is its complexity which leads
to rather involved mathematical expressions when going beyond the simple half-space geometry
that we have considered so far. In fact, this already applies to slab geometries.
Consequently, owing to their more transparent mathematical expressions, we limit the following discussion of thin metal films to the extended hydrodynamic model. To contrast our findings, we report the numerical results for the Boltzmann-Mermin model and provide a physical interpretation in terms of the system's length scales.
The description of a slab's reflection coefficients requires a non-trivial mathematical treatment
when nonlocality is taken into account. The main reason is that the field in a spatially dispersive
medium is not necessarily transverse and the distinction between transverse and longitudinal waves becomes important. For propagation
along the $z$-direction (perpendicular to the surface) in bulk we can associate two field components,
for example $E_x$ and $B_y$, with the transverse propagation and the field component $E_z$ and
the scalar potential $\phi$ with the longitudinal propagation.
For planar geometries such as single- or multi-layered systems stacked in $z$-direction, the
propagation can then be described using the transfer-matrix approach
\cite{yeh1977, yariv1984}.
For a nonlocal layer with thickness $d$ this technique relates the fields at one interface
($z=z_0$) to those on the opposite interface ($z=z_0+d$), i.e.
\begin{eqnarray}\label{eq:4x4tm}
 \begin{pmatrix}
  E_x \\ cB_y \\ E_z \\ \phi
 \end{pmatrix}_{z=z_0+d} ^\sigma
  = \mathbb { Y }^\sigma \left( d \right)
 \begin{pmatrix}
  E_x \\ cB_y \\ E_z \\ \phi
 \end{pmatrix}_{z=z_0}^\sigma,
\end{eqnarray}
where $\mathbb{Y}^\sigma(d)$ is a $4\times 4$ transfer matrix containing information about the
propagation through the medium and $\sigma$ indicates the different transverse polarizations TM and TE. At each
interface, both longitudinal and transverse fields are then connected via ABCs
\cite{Mochan83,Feibelman82,halevi1984, halevi1982, pekar1958}.
The ABCs effectively reduce the four independent fields of \eqref{eq:4x4tm} to the two transverse
fields of the local material adjacent to the nonlocal slab of finite extent,
\begin{equation}
\begin{pmatrix}
E_x \\ cB_y
\end{pmatrix}^\sigma
_ { z = z_0+d} = \mathbb { M } ^ { \sigma }(d)
\begin{pmatrix}
E_x \\ cB_y
\end{pmatrix} _ { z = z_0 }^\sigma.
\end{equation}
The resulting $2\times 2$ transfer matrix is given by
\begin{equation}
\label{eq:2x2trans}
\mathbb{M}^\sigma(d) =
\begin{pmatrix}
\mathrm{cs}_\mathrm{nl}^\sigma(d) & i\delta^\sigma \mathcal{Z}^\sigma_\mathrm{right}(d) \\
i \delta^\sigma/\mathcal{Z}^\sigma_\mathrm{left}(d) & \mathrm{cs}^\sigma_\mathrm{nl}(d)
\end{pmatrix}\,,
\end{equation}
where $\delta^\mathrm{TM/TE}=\pm 1$, while $\mathcal{Z}^\sigma_\mathrm{left}(d)$, $\mathcal{Z}^\sigma_\mathrm{right}(d)$ and
$\mathrm{cs}_\mathrm{nl}^\sigma(d)$ are complex functions describing the propagation in the
nonlocal medium and the interaction with the two interfaces that limit the nonlocal layer.
Here, we refrain from displaying the resulting, rather lengthy expression of
$\mathrm{cs}_\mathrm{nl}^\sigma(d)$, $\mathcal{Z}^\sigma_\mathrm{left}(d)$ and
$\mathcal{Z}^\sigma_\mathrm{right}(d)$. Instead, we refer to the Appendix \ref{sec:AppB} and
to previous works for their detailed description \cite{Intravaia15}.
Applying this transfer-matrix approach, the TM-polarized reflection coefficient for a thin
slab can be written as
\begin{eqnarray}\label{eq:refslab}
r _ { \mathrm { slab } }^\mathrm{TM} &=&
\frac { (Z _ 0^\mathrm{TM}) ^ { 2 } - \mathcal{Z} _ { \mathrm { right } }^\mathrm{TM}  \mathcal{Z} _ { \mathrm { left } }^\mathrm{TM}  } { (Z _ 0^\mathrm{TM}) ^ { 2 } + \mathcal{Z} _ { \mathrm { right } }^\mathrm{TM} \mathcal{Z} _ { \mathrm { left } }^\mathrm{TM}  + 2 \mathrm { i } \operatorname { cs } _ { \mathrm { nl } }^\mathrm{TM}  Z _ 0^\mathrm{TM} \mathcal{Z} _ { \mathrm { left } }^\mathrm{TM} },
\end{eqnarray}
where, in order to shorten the notation, we have dropped the explicit $d$-dependence.
In Fig.~\ref{fig:refk} we depict the imaginary part of the reflection coefficient in
\eqref{eq:refslab} as a function of the wave vector $p$ and compare it to the corresponding result of a half-space geometry
[see \eqref{eq:reflectioncoefficients}]. The graph highlights how in the slab geometry
the behavior for small wavevectors is altered due to the occurrence of the additional
geometric length scale $d$. At low frequencies, for values $p\ll\lambda_\mathrm{TF}^{-1}$
where the material exhibits nonlocal behavior, the imaginary part of the reflection
coefficient in \eqref{eq:refslab} is related to the slab resistivity
\begin{equation}
\label{eq:rhoslab}
 \rho_\mathrm{sb}(p) = \rho_\mathrm{H}(p)\coth[dp] + \rho_\mathrm{Ld}(p).
\end{equation}
Equation~(\ref{eq:rhoslab}) generalizes the expression for the resistivity in \eqref{eq:rhosplit}
to the slab geometry. The two contributions are depicted as gray lines in Fig.~\ref{fig:refk}.
As already shown in Sec.~\ref{sec:halfspace}, the Landau damping dominates the hydrodynamic
resistivity for wave vectors $p\gtrsim \ell^{-1}$. However, for $p\ll 1/d$ the collision-induced
contribution experiences an enhancement via the $\coth[pd]$ factor. The enhancement of the
collision-induced damping in the slab geometry relative to the half-space geometry can be
understood by the symmetric coupling of the two surface plasmon-polaritons (SPPs)
\cite{eguiluz1979, fukui1979, camley1984}
at the two surfaces and has already been discussed for local materials in
\cite{Oelschlager18}.
The transition wave vector, for which the contribution of the enhanced collision-induced and
Landau damping exchange their role, is now given by
$p_\mathrm{sb}\approx\sqrt{\frac{2}{5}} (\omega_\mathrm{p}/\Gamma) \sqrt{d \lambda_\mathrm{TF}}$
and can be obtained by equating the expressions of the corresponding resistivities
[see \eqref{eq:rhoslab}].
In Fig.~\ref{fig:ForceDistanceDep}, we depict the quantum frictional force on an atom moving
above a thin nonlocal slab. For distances $d\ll z_a \ll \ell$, \eqref{eq:fricbulkhydro} is
modified with respect to the half-space results, i.e.
\begin{equation}\label{eq:asympslab}
 \frac{F}{F_\mathrm{lc}} \approx
% \frac{F_\mathrm{H} + F_\mathrm{Ld}}{F_\mathrm{lc}}
% \approx
 \frac{227}{768} \frac{z_a^2}{d^2} + \frac{\omega_\mathrm{p}^2}{\Gamma^2}\frac{73}{240} \frac{\ell^2}{z_a^2}\,.
\end{equation}
The first term coincides with the force expected of a local Drude slab (see Ref. \cite{Oelschlager18}).
It dominates for separations $ z_a\gtrsim \ell_\mathrm{sb} \approx 0.77 \ell \sqrt{d /\lambda_\mathrm{TF}}\sim p_\mathrm{sb}^{-1}$, where the interaction can resolve the structural details, i.e., the finite spatial extension $d$ and the SPP coupling comes into play. The latter enhances the contribution of collision-induced damping and makes the corresponding force less sensitive to changes in the atom-surface separation \cite{Oelschlager18}. Hence, this process dominates for large distances such that the physics is sufficiently described by the local Drude model.
The situation changes when the atom approaches the surface:
The second term of Eq. (\ref{eq:asympslab}) is identically the result of Eq. (\ref{eq:fricbulkhydro}) and corresponds to a bulk response dominated by Landau damping. Below $\ell_{\rm sb}$, the interaction characterized by wavelengths $\lambda\lesssim z_a$ cannot resolve the full geometry of the slab and perceives the surface as an infinitely extended bulk.
For comparison, we report the numerical evaluation of the frictional force using the more involved Boltzmann-Mermin model (see Fig. \ref{fig:ForceDistanceDep}). Noteworthy, the Boltzmann-Mermin model qualitatively and for larger separations (in the near-field description) also quantitatively confirms the findings we already obtained with the extended hydrodynamic description. Since the electron's mean free path that defines the regime of nonlocality, $\ell$, is shorter than the length scale characterizing the transition from bulk to slab physics, $\ell_{\rm sb}$, the model automatically approaches its local limit for separations $z_a\gtrsim\ell_{\rm sb}$.
Consequently, for the chosen material and geometry, the effects of spatial nonlocality play a subleading role as soon as the physics of quantum friction is dominated by the finite slab geometry.
%==================================================================================

% CONCLUSION
%==================================================================================
\section{Conclusion and discussions}\label{sec:concl}
In the present work, we have analyzed the drag force experienced by a moving atom interacting
with a spatially dispersive conductor. We have placed special emphasis on the modeling of the
conductor's low-frequency behavior and have compared an extended hydrodynamic description put
forward by Halevi
\cite{Halevi95}
to the rather involved Boltzmann-Mermin model
\cite{Reiche17}.
The extended hydrodynamic description incorporates a complex-valued frequency-dependent
compressibility factor of the Fermi fluid which can be deduced from an expansion of the
Boltzmann-Mermin model for weak spatial nonlocality. Although much simpler than the
Boltzmann-Mermin model, the extended hydrodynamic description does include dissipative
energy transfer mechanisms in terms of Landau damping and thus sharply contrasts to the
standard hydrodynamic approaches that utilize real-valued and frequency-independent
compressibility factor.
The extended hydrodynamic model is, therefore, well-suited to qualitatively investigate
the impact of nonlocal damping mechanisms on geometrically and/or dynamically complex
quantum-optical setups.

We have illustrated this by computing quantum friction forces for both, infinite half-spaces
and thin metallic slabs. In the half-space geometry, we have shown that the extended
hydrodynamic model essentially contains the enhancement of the force with respect to
a local Drude description predicted by the Boltzmann-Mermin model for atom-surface
separations of the order of the bulk electron's mean free path. This result is related
to Landau damping and is not present in the standard hydrodynamic description. However,
the extended hydrodynamic description overestimates the enhancement by several orders
of magnitude with respect to the more fundamental Boltzmann-Mermin model. On the other
hand, for the slab geometry the extended hydrodynamic description has proved to be very
practical in order to gain reliable insights into the interplay of the different
physical damping processes: For atom-surface separations smaller than $\ell_\mathrm{sb}$,
a characteristic length scale connected to the slab's thickness and the internal damping
mechanisms of the material (see Sec. \ref{sec:slab}), the
interaction is insensitive to the slab's finite thickness, resulting in a frictional
force which equals that obtained for the half-space geometry. For larger separations,
however, the interaction involves the surface-plasmon polaritons which couple across
the slab. The finite thickness of the slab appreciably influences their behavior and the
frictional force experiences a modification in its functional dependence on the atom-surface
separation. Interestingly, as long as the slab thickness is not much smaller than the mean free path, the impact of spatial nonlocality in the
extended hydrodynamic description becomes negligible and the force coincides with the
result using a local Drude description of the material. This phenomenon is connected
to the distinct natures of collision-induced and Landau damping, respectively.
Collision-induced damping describes a multi-scattering average of the quasi-particles
in the Fermi liquid with respect to the wavelength of the electromagnetic mode, which
is also available in spatially local material models independent of the radiation's
wave vector. Landau damping is intrinsically connected to the spatial correlation in
the system and thereby couples non-trivially to the length scales of the slab geometry.
In the slab geometry, the impact of the latter turns out to be more constrained with respect to the
half-space geometry.

The standard hydrodynamic model has become popular in describing the nonlocal optical
 response of nano-photonic systems
\cite{Ciraci13,raza13}.
Commonly, one has to solve Maxwell's equations in non-trivial geometries and relies
here on sophisticated numerical solvers (see e.g. Ref.
\cite{Busch11}
for a method in the time-domain), where material descriptions are included by means
of partial differential equations for macroscopic electrodynamic quantities fulfilling
additional boundary conditions
\cite{raza13}.
Including descriptions as complex as the Boltzmann-Mermin model is, therefore, numerically
a highly non-trivial problem (especially for non-planar geometries) and researchers are pushing towards extensions of the
hydrodynamic equations
\cite{Mortensen14,Raza15}.
Our analyses delineates the limits of applicability of the aforementioned material
models in the context of nonequilibrium fluctuation-induced interactions and provides
a transparent physical interpretation of the roles of collision-induced and Landau
damping. In view of the serious challenges related to a numerical implementation of
the more fundamental Boltzmann-Mermin model and the relative ease of implementing
the standard hydrodynamic model, our work is also of relevance for further investigations regarding extensions of the hydrodynamic description.
An example is provided by the analysis of Casimir(-Polder) forces in systems involving spatially nonlocal material and non-trivial geometries beyond the analytically accessible cases. \cite{esquivel04,reiche18arxiv}.%==================================================================================

% ACKNOWLEDGEMENTS
%==================================================================================
\section*{Acknowledgements}
We thank D. Huynh for fruitful discussions.
We acknowledge support from the Deutsche Forschungsgemeinschaft (DFG) through the DIP program (Grant No. SCHM 1049/7-1) and the
DFG Collaborative Research Center (CRC) 951 Hybrid Inorganic/ Organic Systems for Opto- Electronics (Project B10).

%%%%%%%%%%%%%%%%%%%%%%% APPENDIX %%%%%%%%%%%%%%%%%%%%%%%%%%%
\appendix

\section{Resistivity and friction using the hydrodynamic model}
\setcounter{equation}{0}
\renewcommand{\theequation}{A{\arabic{equation}}}
\label{appendix1}
In this Appendix, we derive the expression for the resistivity of the hydrodynamic
material model, Eqs. (\ref{eq:rhononlocal}) and (\ref{eq:rhosplit}), and deduce the
quantum frictional force experienced by an atom in motion relative to a material-filled
half-space as reported in Eq. (\ref{eq:fricbulkhydro}).
Starting from Eq. (\ref{eq:surfaceimpedanceP}) and inserting the relation for the
permittivities given by Eq. (\ref{eq:permittivitiesH}), we obtain for the TM surface
impedance of the hydrodynamic model
\begin{align}
\frac{Z^{\rm TM}}{Z^{\rm TM}_0}=\frac{1}{\epsilon_{\rm Dr}\kappa}\left(\kappa_{\epsilon}+\frac{p(\epsilon_{\rm Dr}-1)}{\sqrt{1+\frac{\omega_\mathrm{p}^2}{\beta^2p^2}\frac{\epsilon_{\rm Dr}}{\epsilon_{\rm Dr}-1}}}\right),
\end{align}
where we have defined $\kappa=\sqrt{p^2-\omega^2/c^2}$ and
$\kappa_{\epsilon}=\sqrt{p^2-\epsilon_{\rm Dr}\omega^2/c^2}$ with
$\text{Im}\kappa,\text{Im}\kappa_{\epsilon}<0$.
In the near-field regime (formally $c\to\infty$), the TM reflection coefficient
reduces to
\begin{align}
r^{\rm TM}=\frac{\epsilon_{\rm Dr}-1}{\epsilon_{\rm Dr}+1+2(\epsilon_{\rm Dr}-1)\frac{\beta^2p^2}{\omega_\mathrm{p}^2}\left(1+\sqrt{1+\frac{\omega_\mathrm{p}^2}{\beta^2 p^2}\frac{\epsilon_{\rm Dr}}{\epsilon_{\rm Dr}-1}}\right)}.
\end{align}
In the limit of vanishing spatial dispersion (formally $\beta\to 0$), this last
result reduces to the usual local expression $r^{\rm TM}\to (\epsilon_{\rm Dr}-1)/(\epsilon_{\rm Dr}+1)$.
For frequencies $\omega\ll\Gamma$, using the frequency dependent compressibility factor $\beta(\omega)$ given in \eqref{eq:betaHalevi} the reflection coefficient behaves Ohmic (see
main text), i.e.
\begin{align}
r^{\rm TM}\approx R(\lambda_{\rm TF}p)+2\imath\epsilon_0\rho_{\rm eH}(\lambda_{\rm TF}p)\omega,
\end{align}
where we have defined the real function
\begin{align}
R(x)&=1+2x\left(x-\sqrt{1+x^2}\right)\to
\begin{cases}
1, &x\ll 1\\
(2x)^{-2}, &x\gg 1
\end{cases},
\end{align}
which corresponds to the reflection coefficient for $\omega=0$ and exhibits an absolute value that is smaller or equal to one. Further, the resistivity
decomposes into a sum of two terms, $\rho_{\rm eH}(\lambda_{\rm TF}p)=\rho_{\rm H}(\lambda_{\rm TF}p)+\rho_{\rm Ld}(\lambda_{\rm TF}p)$, associated with distinct damping mechanisms. The first one is connected to
collision-induced damping of quasi-particles in the Fermi liquid,
\begin{align}
\rho_{\rm H}(x)&=\frac{\Gamma}{\epsilon_0\omega_\mathrm{p}^2}\frac{  \frac{x}{2} \left(\sqrt{x^2+1}+2 x\right)+1
}{\left(x^2+1\right) \left(2 x \left(\sqrt{x^2+1}+x\right)+1\right)^2 }
\nonumber\\
&\to\frac{\Gamma}{\epsilon_0\omega_\mathrm{p}^2}
\begin{cases}
1 ,&x\ll1\\
\frac{3}{32x^2}\frac{1}{2+x^2} ,&x\gg1
\end{cases}
\end{align}
and the other is connected to Landau damping in the bulk,
\begin{align}
\rho_{\rm Ld}(x)&=
\frac{\imath}{\epsilon_0}
\frac{\beta'(0)}{\beta(0)}
\frac{ x \left(2 x^2 \left(\sqrt{x^2+1}+x\right)+\sqrt{x^2+1}+2x\right) }{\left(x^2+1\right) \left(2 x \left(\sqrt{x^2+1}+x\right)+1\right)^2 }
\nonumber\\
&\to \frac{\imath}{\epsilon_0}
\frac{\beta'(0)}{\beta(0)}
\begin{cases}
x ,&x\ll1\\
\frac{1}{4}\frac{1}{1+x^2} ,&x\gg1
\end{cases}.
\end{align}
Here, the prime denotes the derivative with respect to frequency. This second term only appear
for a complex-valued and frequency-dependent compressibility factor $\beta(\omega)$ and in the main text
this was associated with a resistivity originating from Landau damping. Models with real-valued
$\beta$ only feature the collision-induced resistivity $\rho_{\rm H}$.
Further, while both $R(x)$ and $\rho_{\rm H}$ are monotonously
decreasing functions of $x$, $\rho_{\rm Ld}(x)$ first monotonously increases, then
exhibits a maximum at $x=\sqrt{\sqrt{2}^{-1}-2^{-1}}\approx 0.46$, and finally monotonously
approaches zero for increasing $x$. This behavior becomes apparent in Fig. \ref{fig:rho}.
From Eq. (\ref{eq:generalAsymp}), we infer that the low-velocity quantum frictional force
is given in terms of the function $\mathcal{D}_n(z_a)$ with $n=0,1,2$. Since the latter is
a linear functional of the resistivity, it also decomposes into two contributions related
to collision-induced and Landau damping, respectively. In particular, we have
\begin{align}
\mathcal{D}^{\rm eH}_n(z_a)=\mathcal{D}_n^{\rm H}(z_a)+\mathcal{D}_n^{\rm Ld}(z_a),
\end{align}
which can be evaluated exactly for the limiting cases of small and large $x=p\lambda_{\rm TF}$,
respectively, yielding
\begin{subequations}\label{DAsymp}
\begin{align}
&\mathcal{D}_{\{0,1,2\}}^{\rm H}\\\nonumber
&\to
	2\epsilon_0\rho_{\rm lc}
	\begin{cases}
	\left\{\frac{3\pi}{64\sqrt{2}\lambda_{\rm TF}^3}
	      , \frac{1}{z_a}\frac{3}{64\lambda_{\rm TF}^4}
	      , \frac{1}{z_a^3}\frac{3}{128\lambda_{\rm TF}^4}\right\}
	, & z_a\ll\lambda_{\rm TF}\\
	\left\{\frac{1}{4z_a^3}
		  , \frac{3}{4z_a^5}
		  , \frac{45}{8z_a^7}\right\}
	, & z_a\gg\lambda_{\rm TF}
	\end{cases}
	\\
&\mathcal{D}_{\{0,1,2\}}^{\rm Ld}\\\nonumber
&\to
	\frac{4}{5\Gamma}
	\begin{cases}
	\left\{\frac{1}{z_a}\frac{1}{8\lambda_{\rm TF}^2}
	      , \frac{1}{z_a^3}\frac{1}{16\lambda_{\rm TF}^2}
	      , \frac{1}{z_a^5}\frac{3}{16\lambda_{\rm TF}^2}\right\}
	, & z_a\ll\lambda_{\rm TF}\\
	\left\{\frac{1}{z_a^4}\frac{3\lambda_{\rm TF}}{8}
	      , \frac{1}{z_a^6}\frac{15\lambda_{\rm TF}}{8}
	      , \frac{1}{z_a^8}\frac{315\lambda_{\rm TF}}{16}\right\}
	, & z_a\gg\lambda_{\rm TF}
	\end{cases}.
\end{align}
\end{subequations}
Here, we have introduced the abbreviation $\rho_{\rm lc}=\Gamma/(\epsilon_0\omega_\mathrm{p}^2)$
and have used that $\imath\beta'(0)/(\epsilon_0\beta(0))=2/(5\epsilon_0\Gamma)$. For
typical metals, the Thomas-Fermi screening length $\lambda_{\rm TF}$ lies in the range
of a few angstroms, a range for that our theory is not valid anymore.
Hence, for all practical cases, we have the situation that $z_a\gg\lambda_{\rm TF}$ and
the second set of lines of Eqs. (\ref{DAsymp}) apply.
This means that $\mathcal{D}_n^{\rm H}\approx \mathcal{D}_n^{\rm lc}$ reduces to the
local Drude result $\mathcal{D}_n^{\rm lc}$ reported in Ref.
\cite{Intravaia16}
and we obtain for the quantum frictional force experienced by an atom in the half-space
geometry
\begin{align}
\frac{F}{F_{\rm lc}}-1&\approx
\frac{\frac{\Phi_0\Phi_2}{3}\left(\mathcal{D}_0^{\rm lc}\mathcal{D}_2^{\rm Ld}+\mathcal{D}_0^{\rm Ld}\mathcal{D}_2^{\rm lc}+\mathcal{D}_0^{\rm Ld}\mathcal{D}_2^{\rm Ld}\right)}{\frac{\Phi_0\Phi_2}{3}\mathcal{D}_0^{\rm lc}\mathcal{D}_2^{\rm lc}+\Phi_1^2\left[\mathcal{D}_1^{\rm lc}\right]^2}
\nonumber\\
&
+\frac{\Phi_1^2\left(2\mathcal{D}_1^{\rm lc}\mathcal{D}_1^{\rm Ld}+\left[\mathcal{D}_1^{\rm Ld}\right]^2\right)}{\frac{\Phi_0\Phi_2}{3}\mathcal{D}_0^{\rm lc}\mathcal{D}_2^{\rm lc}+\Phi_1^2\left[\mathcal{D}_1^{\rm lc}\right]^2},
\end{align}
where $F_{\rm lc}$ is the local Drude result (see main text). Expressing the atomic dipole
moment $\mathbf{d}$ in polar coordinates and averaging over all possible directions, we
obtain
\begin{equation}
\langle\frac{\Phi_0\Phi_2}{3}\rangle=\frac{7}{7680\pi^2}\frac{\alpha_0^2}{\epsilon_0^2}\quad
\langle\Phi_1^2\rangle=\frac{29}{15360\pi^2}\frac{\alpha_0^2}{\epsilon_0^2}
\end{equation}
where $\alpha_{0}=\sum_{i}\alpha_{ii}/3$.
This leads to
\begin{align}
\frac{F}{F_{\rm lc}}\approx1+\frac{\omega_\mathrm{p}^2}{\Gamma^2}\left(2\frac{\lambda_{\rm TF}}{z_a}+\frac{73}{240}\frac{\ell^2}{z_a^2}\right).
\end{align}
Noticing that the second term is dominant in the nonlocal regime, the expression restores the relation
reported in \eqref{eq:fricbulkhydro}. From the previous equation, we conclude that the frictional
force using the extended hydrodynamic model differs substantially from the local Drude
result $F_{\rm lc}$ for separations
\begin{align}
z_a\ll \bar{\ell}:= \left(\frac{1}{\sqrt{3}}+\frac{1}{4}\sqrt{\frac{51}{5}}\right)\frac{\omega_\mathrm{p}}{\Gamma}\ell\approx 1.4\frac{\omega_\mathrm{p}}{\Gamma}\ell.
\end{align}
This has been confirmed by numerical evaluation (see Fig. \ref{fig:ForceDistanceDep}).

\section{Transfer-Matrix Approach using the hydrodynamic model}
\setcounter{equation}{0}
\renewcommand{\theequation}{B{\arabic{equation}}}
\label{sec:AppB}
As discussed in Sec.~\ref{sec:slab}, without ABCs the hydrodynamic model yields four linearly
independent variables $\left(E_x,\,cB_y,\,E_{z},\,\phi\right)$
\cite{Mochan83}.
In the transfer-matrix approach, the propagation of those fields through a single
hydrodynamic slab is written in terms of a $4\times4$ matrix $\mathbb{Y}(d)^\sigma$ with
respect to these four variables [see \eqref{eq:4x4tm}]. This matrix,
\begin{equation}\label{eq:4x4app}
\mathbb{Y}(d)^\sigma
=
\mathbb{Z}_\mathrm{H}^\sigma \mathbb{P}_\mathrm{H}(d)\left(\mathbb{Z}_\mathrm{H}^\sigma\right)^{-1},
\end{equation}
is composed of a propagation matrix of left- and right-propagating transverse and
longitudinal fields
\begin{equation}
\mathbb{P}_\mathrm{H}(d) = \mathrm{diag}\left[
e^{iq_t d}, e^{-iq_l d}, e^{iq_l d}, e^{-iq_l d},
\right]
\end{equation}
where $q_{t/l}$ are implicitly defined via
\begin{equation}
\epsilon_t(\omega)\omega^2=(p^2+q_t^2)c^2 \quad\mathrm{and}\quad
\epsilon_l(\omega,\sqrt{p^2+q_l^2}) = 0,
\end{equation}
which is ``sandwiched'' between two surface impedance matrices at the respective
interfaces. In fact, owing to the symmetry of the system, one of the interface
matrices is the inverse of the other [as anticipated in
\eqref{eq:4x4app}]. Explicitly, the interface matrix reads
\begin{align}
\mathbb{Z}_\mathrm{H}^\sigma&=-\frac{1}{2\epsilon_0 c}
\begin{pmatrix}
Z_\mathrm{Dr}^\sigma & Z_\mathrm{Dr}^\sigma & ik & ik \\
1 & -1 & 0 & 0 \\
-W_\mathrm{Dr} & W_\mathrm{Dr} & iq_l & -iq_l \\
0&0&-1&-1
\end{pmatrix},
\end{align}
with
\begin{equation}
Z _ { \mathrm { Dr } } ^ \mathrm{ TM } = \frac { c q _ { \mathrm { t } } } { \omega \epsilon _ { \mathrm { Dr } } ( \omega ) } , \quad Z _ { \mathrm { Dr } } ^ \mathrm{ TE } = \frac { \omega } { c q _ { \mathrm { t } } } , \quad W _ { \mathrm { Dr } } = \frac { c k } { \omega \epsilon _ \mathrm{ Dr } ( \omega ) }.
\end{equation}
However, if we apply the ABCs at the interfaces, we match the fields of the nonlocal
material with the fields of the adjacent local material and thus obtain the reduced
$2\times 2$ transfer matrix for local materials, as written in \eqref{eq:2x2trans}
of the Sec.~\ref{sec:slab}. This matrix relates to $\mathbb{Y}$ of \eqref{eq:4x4app}
as follows
\begin{align}
\mathbb { M }  _ { 11 } &= \mathbb { Y } _ { 11 } - \mathbb { Y } _ { 14 } \frac { W _ { 0 } \mathbb { Y } _ { 21 } + \mathbb { Y } _ { 31 } } { W _ { 0 } \mathbb { Y } _ { 24 } + \mathbb { Y } _ { 34 } }
\\
\mathbb { M }_ { 12 } &= \mathbb { Y } _ { 12 } - W _ { 0 } \mathbb { Y } _ { 13 }  \nonumber\\
&\quad -\mathbb { Y } _ { 14 } \left( \frac { W _ { 0 } \mathbb { Y } _ { 22 } + \mathbb { Y } _ { 32 } } { W _ { 0 } \mathbb { Y } _ { 24 } + \mathbb { Y } _ { 34 } } - W _ { 0 } \frac { W _ { 0 } \mathbb { Y } _ { 23 } + \mathbb { Y } _ { 33 } } { W _ { 0 } \mathbb { Y } _ { 24 } + \mathbb { Y } _ { 34 } } \right)
\nonumber\\
\mathbb { M } _ { 21 } &= \mathbb { Y } _ { 21 } - \mathbb { Y } _ { 24 } \frac { W _ { 0 } \mathbb { Y } _ { 21 } + \mathbb { Y } _ { 31 } } { W _ { 0 } \mathbb { Y } _ { 24 } + \mathbb { Y } _ { 34 } }
\nonumber\\
\mathbb { M } _ { 22 } &= \mathbb { Y } _ { 22 } - W _ { 0 } \mathbb { Y } _ { 23 } \nonumber\\
&\quad- \mathbb { Y } _ { 24 } \left( \frac { W _ { 0 } \mathbb { Y } _ { 22 } + \mathbb { Y } _ { 32 } } { W _ { 0 } \mathbb { Y } _ { 24 } + \mathbb { Y } _ { 34 } } - W _ { 0 } \frac { W _ { 0 } \mathbb { Y } _ { 23 } + \mathbb { Y } _ { 33 } } { W _ { 0 } \mathbb { Y } _ { 24 } + \mathbb { Y } _ { 34 } } \right),
\nonumber
\end{align}
where $W_0=\frac{ck}{\omega}$. In the previous equation, we have dropped the polarization
index $\sigma$. For further details, we refer to Ref.~\cite{Intravaia15}.

%%%%%%%%%%%%%%%%%%%%%%% References %%%%%%%%%%%%%%%%%%%%%%%%%

%\bibliography{./bib/extracted}

\begin{thebibliography}{10}
\bibitem{footnote1}
In the idealized case of a perfect crystal, collision-induced damping
  approaches zero with vanishing temperature (see Ref. \cite{hwang19} for a
  recent discussion). This phenomenon generated discussions about the behavior
  of the equilibrium Casimir interaction and its thermodynamical consistency
  (see Refs. \cite{svetovoy05,svetovoy08} and \cite{reiche18arxiv} for recent
  treatments in the context of Casimir and Casimir-Polder interactions,
  respectively). However, in the present description of quantum friction, the
  impact of the transverse electric reflection coefficient turns out subleading
  especially at short atom-surface separations (near-fied limit)
  \cite{Intravaia16a}.
\end{thebibliography}
%\bibliographystyle{/media/daniel/data/Documents/Promotion/bib/bibstyle/prstytitlenew}

\end{document}